\documentclass{Interspeech}
\usepackage{xcolor}
\usepackage{multirow}
\usepackage{booktabs}

\newcommand\ourmethod{FlowTSE}

\interspeechcameraready

\title{FlowTSE: Target Speaker Extraction with Flow Matching}

\author[]{Aviv}{Navon}
\author[]{Aviv}{Shamsian}
\author[]{Yael}{Segal-Feldman}
\author[]{Neta}{Glazer}
\author[]{Gil}{Hetz}
\author[]{Joseph}{Keshet}

\affiliation{}{aiOla Research}{Israel}
\email{\{aviv, aviv.shamsian\}@aiola.com}


\usepackage{comment}

\begin{document}

\maketitle

\begin{abstract}
    
    Target speaker extraction (TSE) aims to isolate a specific speaker's speech from a mixture using speaker enrollment as a reference. While most existing approaches are discriminative, recent generative methods for TSE achieve strong results. However, generative methods for TSE remain underexplored, with most existing approaches relying on complex pipelines and pretrained components, leading to computational overhead. In this work, we present FlowTSE, a simple yet effective TSE approach based on conditional flow matching. Our model receives an enrollment audio sample and a mixed speech signal, both represented as mel-spectrograms, with the objective of extracting the target speaker's clean speech. Furthermore, for tasks where phase reconstruction is crucial, we propose a novel vocoder conditioned on the complex STFT of the mixed signal, enabling improved phase estimation. Experimental results on standard TSE benchmarks show that FlowTSE matches or outperforms strong baselines.\footnote{Audio samples are available at:\\ \textcolor{magenta}{\url{https://aiola-lab.github.io/flow-tse}}}
    

\end{abstract}

\section{Introduction}

Target Speaker Extraction (TSE) focuses on extracting the clean speech of a specific speaker from an audio mixture containing interfering speakers and background noise. TSE has gained significant attention for its role in improving speech clarity in communication systems and enhancing hearing aids for individuals with hearing impairments~\cite{borsdorf2024wtimit2mix,zhang2024ddtse}. Additionally, TSE enhances Automatic Speech Recognition (ASR) by isolating the target speaker's speech before transcription, reducing interference from noise and other speakers~\cite{polok2024target}. This enables applications like virtual assistants to operate accurately in noisy environments.
Traditional approaches to TSE mainly rely on discriminative models, which directly map a noisy speech sample to a clean speech or speech mask, by optimizing signal-level metrics such as signal-to-distortion ratio~\cite{wang2018supervised,fu2018end}. While these models have shown strong performance, they often produce artifacts and may fail to reconstruct highly corrupted samples~\cite{yu2024generation}. In addition, they may exhibit limited generalization to unseen noise types or speakers \cite{zhang2024ddtse,richter2023speech,lemercier2023storm,lemercier2023analysing}.

\begin{figure}[t]
\centering
\includegraphics[width=0.555\linewidth]{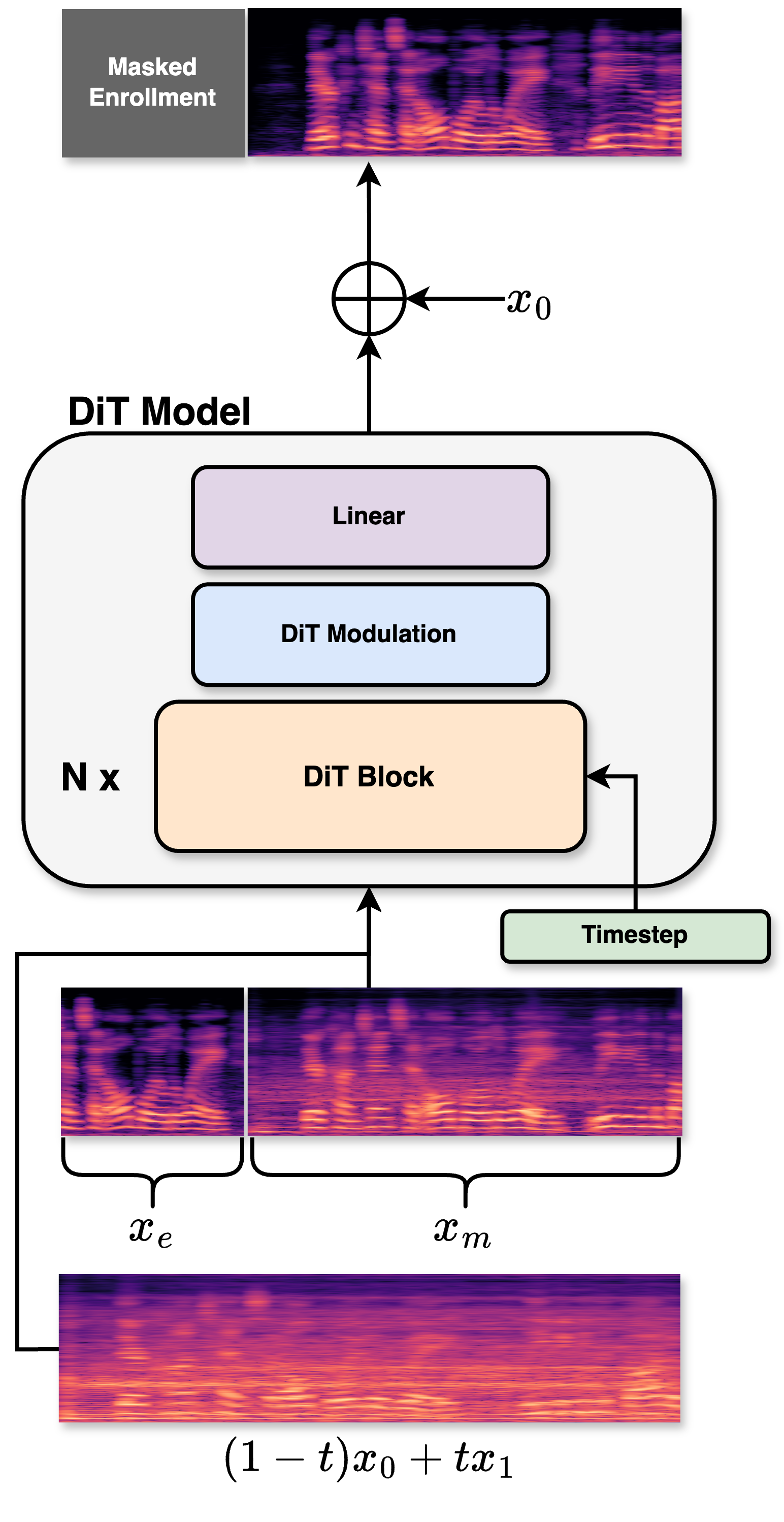}
\caption{\textit{\ourmethod{} architecture. The model is conditioned on the mixed and enrollment mel.}}
\label{fig:arch}
\end{figure}

Recently, generative models have emerged as a powerful alternative, especially in speech enhancement, offering improved perceptual quality and robustness~\cite{richter2023speech,lemercier2023storm,lu2022conditional}. Despite these advantages, their potential in TSE is still relatively unexplored. In the context of TSE, generative approaches model the distribution of clean speech given a target speaker’s reference, enabling more natural and high-fidelity speech extraction \cite{zhang2024ddtse,yu2024generation}. However, current generative models also have their limitations. Many existing methods rely on complex pipelines involving discriminative TSE models~\cite{zhang2024ddtse}, or incorporating pretrained models such as Self-Supervised Learning (SSL) models \cite{yu2024generation,tang2024tselm} or ASRs~\cite{ma2025enhancing}, which complicates their training and deployment. Additionally, several approaches operate directly in the complex short-time Fourier transform (STFT) domain, which, while providing fine-grained frequency resolution, can introduce computational overhead, potential artifacts, and slow convergence \cite{zhang2024ddtse,kamo2023target}.

An alternative is to model speech in the mel-spectrogram domain, simplifying the pipeline and reducing computational complexity. However, this approach comes with its own challenges, particularly phase reconstruction. Since mel-spectrograms discard phase information, vocoder-based inversion may introduce artifacts or degrade speech quality, especially in scenarios where precise phase reconstruction is crucial. Moreover, generative models may still struggle with maintaining speaker identity in challenging conditions \cite{ma2025enhancing}.

To address these challenges, we propose a simplified generative approach based on Flow Matching (FM) \cite{lipman2022flow} operating directly on mel-spectrograms. This representation reduces training complexity while maintaining high perceptual quality. Furthermore, for applications where phase reconstruction is crucial, we propose conditioning the vocoder on the comple STFT of the mixed signal, supporting better phase estimation. 

We conduct extensive evaluations on both clean two-speaker mixtures and more challenging two-speaker scenarios with background noise. Our results demonstrate that \ourmethod{} achieves performance matches or outperforms strong baselines while maintaining a simpler and more efficient pipeline.

Our main contributions are as follows: (i) We propose \ourmethod{}, a novel TSE approach based on conditional flow matching, offering a simple yet effective alternative to complex generative pipelines. 
(ii) We introduce a vocoder augmentation strategy to improve phase reconstruction in scenarios where phase estimation is critical. (iii) We provide extensive evaluations in both clean and noisy TSE conditions, demonstrating the effectiveness and robustness of our approach.

\section{Related Work}


TSE falls under the broader category of Speech Enhancement (SE), which aims to recover clean speech from an input audio corrupted by noise. SE methods are generally categorized as discriminative or generative. Discriminative approaches directly map noisy inputs to clean outputs, operating either in the frequency domain \cite{wang2018supervised} or in the time domain to mitigate phase estimation distortions \cite{fu2018end}. Generative approaches enhance robustness and perceptual quality, with recent advances incorporating diffusion models for both time-domain \cite{lu2022conditional} and complex STFT-domain processing \cite{richter2023speech, lemercier2023storm}. Additionally, hybrid models, such as \cite{kim2024guided}, integrate discriminative and generative techniques to achieve improved performance.

Similar to SE, TSE methods can be categorized into discriminative and generative approaches. Discriminative models either operate in the frequency domain, such as \cite{wang2018voicefilter}, or in the time domain, as seen in \cite{ge2020spex}. However, these methods often struggle to generalize to unseen noise conditions or speakers \cite{richter2023speech, zhang2024ddtse, lemercier2023storm}, leading to increased interest in generative methods. In order to address these limitations, generative-based TSE models have emerged. \cite{kamo2023target} proposed a diffusion model operating on the complex STFT, while \cite{zhang2024ddtse} combined diffusion processes with discriminative objectives. \cite{ma2025enhancing} further improved generative models by incorporating speech embeddings and ASR loss, enhancing both perceptual quality and intelligibility. Another approach is to leverages discrete speech tokens derived from large self-supervised language models, as demonstrated by \cite{yu2024generation} and \cite{tang2024tselm}.
However, there is still room for improvement, as the state-of-art methods rely on complex pipelines involving discriminative TSE models~\cite{zhang2024ddtse} or large pretrained models~\cite{yu2024generation,tang2024tselm, ma2025enhancing}.

\begin{table*}[t]
    \centering
    \caption{Performance comparison on Libri2Mix (clean and noisy) using the \texttt{min} configuration. We bold the best results and underline the second-best results.}
    \resizebox{\textwidth}{!}{
    \begin{tabular}{l c ccccc c ccccc}
        \toprule
        & & \multicolumn{5}{c}{Libri2Mix Noisy (mix-both)} & & \multicolumn{5}{c}{Libri2Mix Clean (mix-clean)} \\
        \cmidrule{3-7} \cmidrule{9-13}
        & Type & PESQ & ESTOI & OVRL & DNSMOS & SIM & & PESQ & ESTOI & OVRL & DNSMOS & SIM  \\
        \midrule
        Mixture  &  -  & 1.08  & 0.40  & 1.63  & 2.71  & 0.46  && 1.15  & 0.54  & 2.65  & 3.41  & 0.54 \\
        Clean & - & - & - & 3.26 & 3.73 & - && - & - & 3.26 & 3.73 & - \\
        \midrule
         DPCCN     & D  & \underline{1.74}  & 0.73  & 2.93 & 3.58  & 0.69  & & 2.22  & 0.83  & 3.05  & 3.73  & 0.82 \\
       NCSN++     & D & 1.55  & 0.73  & 3.15  & 3.68  & 0.69  & & \underline{2.24}  & \textbf{0.86}  & \underline{3.28}  & \textbf{3.86}  & \underline{0.85} \\
         \midrule
        DiffSep+SV & G & 1.32  & 0.60  & 2.78  & 3.63  & 0.62  & & 1.85 & 0.79  & 3.14  & \underline{3.83}  & 0.83 \\
        DDTSE  & G &  1.60  & \underline{0.71}  & \underline{3.28}  & \underline{3.74}  & \underline{0.71}  & & 1.79  & 0.78  & \textbf{3.30}  & 3.79  & 0.73 \\
        
        SKIM & G &  1.29 & /  & 3.27  & /  & /  & &  / &  / & / & /  & / \\
        \midrule
        \ourmethod{} & G & \textbf{1.86} & \textbf{0.75} &  \textbf{3.30} & \textbf{3.82}  & \textbf{0.83} && \textbf{2.58} & \underline{0.84} & 3.27 & 3.79 & \textbf{0.90} \\
        \bottomrule
    \end{tabular}
}
    \label{tab:libri2mix}
\end{table*}

\section{Method}\label{sec:method}

In this section, we present \ourmethod{}, our flow matching-based approach for TSE. Our method maintains a simple pipeline that does not rely on external pretrained models besides the vocoder, which converts mel spectrograms into audio. Additionally, we propose a simple yet effective modification to vocoder models, specifically Vocos~\cite{siuzdak2023vocos}, to enhance phase reconstruction in scenarios where precise phase reconstruction is essential.

\subsection{Preliminaries: Flow Matching}

Flow matching (FM) ~\cite{lipman2022flow} models a probability path $(p_t)_{0 \leq t \leq 1}$ between a known source distribution  $p_0 = p$ and the target data distribution $p_1 = q$. Instead of learning a score function as in diffusion models, FM directly models a time-dependent velocity field \( u_t: [0,1] \times \mathbb{R}^d \to \mathbb{R}^d \), parameterized by a neural network. This velocity field defines a transformation \( \psi_t: [0,1] \times \mathbb{R}^d \to \mathbb{R}^d \) satisfying the ordinary differential equation (ODE),
$$\frac{d}{dt} \psi_t(x) = u_t(\psi_t(x)),$$
where \( \psi_t(x) \) represents the flow of a sample along the probability path. Training FM involves regressing the predicted velocity field to its true counterpart. In practice, we replace the FM objective with the conditional flow matching (CFM) objective. Here, a target sample \( x_1 \sim q \) is paired with a source sample \( x_0 \sim p \), and the velocity field is learned to match the conditional velocity:
$$\mathcal{L}_{\text{CFM}}(\theta) = \mathbb{E}_{t, x_t, x_1} \lVert u_t^\theta (x_t) - u_t(x_t | x_1) \rVert^2,$$
with $x_t:=\psi_t(x_0)$ and where $\theta$ are trainable parameters of a neural network. The CFM and FM objectives were proven to have identical gradients w.r.t $\theta$~\cite{lipman2022flow}. 
Furthermore, under an optimal transport setting, this simplifies to:
\begin{equation}\label{eq:cfm}
    \mathcal{L}_{\text{CFM}}(\theta) = \mathbb{E}_{t, x_0, x_1} \lVert u_t^\theta (x_t) - (x_1 - x_0) \rVert^2,
\end{equation}
where $x_t = (1 - t)x_0 + t x_1$. At inference, starting from  $x_0 \sim p_0$, we integrate the learned velocity field $u_t^{\theta}$ along the probability path using numerical solvers, to obtain $x_1 \sim p_1 $ from the source distribution. Also, classifier-free guidance can be applied to enable better control over the generation process.

\subsection{FlowTSE}


We aim to design a generative approach for TSE with a simplified pipeline. Motivated by recent advancements in text-to-speech, our model adopts a similar architecture to~\cite{chen2024f5}, utilizing diffusion transformer blocks~\cite{peebles2023scalable} and adaptive layer normalization for processing flow time step conditioning $t$. Different from~\cite{chen2024f5}, we omit all text-related inputs and modules, focusing solely on audio-based conditioning. The input consists of an enrollment audio sample $x_e$ and a mixed speech signal $x_m$, both represented as mel-spectrograms. These are concatenated along the time dimension to form the model input conditioning $x=[x_e, x_m]\in \mathbb{R}^{(t_e+t_m)\times d}$, where $d$ is the number of mel channels, and $t_e, t_m$ denote the time dimension of the enrollment and mixed speech, respectively. 
During self-attention, we allow mixed tokens to attend to the enrollment tokens, but not the other way around.
The target output $y\in\mathbb{R}^{t_m\times d}$ is the clean mel of the desired speaker. The \ourmethod{} architecture is depicted in Figure~\ref{fig:arch}.

\textbf{Training.} During training, given a data point $(x_e, x_m, y)$, we sample $t\sim U[0, 1]$ and set $x_t=(1-t)x_0+ty$, where $x_0$ is sampled from the source distribution $p_0=N(0,I)$. The model is trained using the CFM objective in Eq.~\ref{eq:cfm}.

\textbf{Inference.} At inference, given an enrollment $x_e$ and a mixed signal $x_m$, we sample $x_0\sim p_0$ and use an ODE solver to integrate the train velocity field $u_t^{\theta}(x_t)$ from $t{=}0$ to $t{=}1$. The enrollment part of the generated mel is omitted and a vocoder is utilized to convert the mel into a waveform. In the following section, we propose an approach to modify the Vocoder to achieve accurate phase reconstruction by conditioning on complex STFT of the mixed signal. 

\subsection{Phase-conditioned Vocoder}\label{sec:vocoder}

One key challenge for generative-based TSE approaches utilizing mel-based vocoders is the lack of explicit phase modeling, as mel-spectrograms do not retain phase information. While a simple recontract phase is sufficient for most applications, precise phase estimation may be necessary when our TSE method is integrated with systems that operate directly on the waveform, such as HuBERT \cite{hubert}, or when used in hearing aids and professional audio processing, for which accurate waveform reconstruction is crucial.
We extend the architecture of the Vocos vocoder~\cite{siuzdak2023vocos} by introducing phase-aware capabilities, specifically designed for the TSE task. The key modifications are as follows. First, we augment the ConvNeXt~\cite{liu2022convnet} blocks with cross-attention layers that condition the mel-spectrogram processing on the mixed speech STFT features. This enables the model to leverage phase information from the mixed signal. 
Additionally, we modify the standard iSTFT Vocos head to allow explicit phase modeling through learnable complex coefficients. This head predicts both magnitude and phase components, and combines them with the mixed signal's phase information through a complex-valued linear combination $(\alpha \odot s_m + \beta \odot s_p)$, where $\alpha$ and $\beta$ are learned for each time-frequency bin, $s_m$ is the mixed STFT and $s_p$ is the predicted STFT. Here, $\odot$ denotes the Hadamard product. This modification of the standard iSTFT is motivated by the alignment approach proposed in~\cite{lutati2024separate}.
The new architecture transform Vocos from a general-purpose neural vocoder into a phase-aware vocoder suited for TSE. Importantly, this component can be trained independently, using the mel of the clean speech and complex STFT of the mixed signal as inputs. We optimize the vocoder using the SI-SDR objective. Evaluation of this approach is presented in Section \ref{sec:vocoder_exp}.

\section{Experiments}

\subsection{Experimental Setup}

\textbf{Datasets.} We follow~\cite{ma2025enhancing}  to train our model using the LibriSpeech dataset~\cite{panayotov2015librispeech}. Specifically, we use the \textnormal{\texttt{train clean 100}} and \textnormal{\texttt{train clean 360}} subsets. During training, we sample two random samples together with a noise sample from the WHAM!~\cite{Wichern2019WHAM} dataset. The speech utterances are mixed with an SNR ratio sampled from $U[-5, 5]$ dB. The noise sample is used with probability $0.75$, and in that case, it is merged with SNR sampled from $U[-5, 5]$ dB. The enrollment is a speech segment of the target speaker, extracted from a different utterance, with a duration randomly sampled between 1 and 5 seconds.
For TSE evaluation, we use two subsets of the Libri2Mix~\cite{cosentino2020librimix} dataset, \texttt{mix-both} and \texttt{mix-clean} under the \texttt{min} and  \texttt{max} settings. The \texttt{mix-both} is constructed from merging two speech utterances and noise sampled from the WHAM! dataset. Similarly to training, an enrollment of up to 5 seconds is sampled from a different utterance of the target speaker. For SE evaluation, we use the Libri2Mix \texttt{mix-single} subset with the \texttt{min} settings, created by merging a single speech utterance and noise sampled from the WHAM! dataset.

\textbf{Baselines.} We evaluate our model against several TSE baselines. Generative-based methods include: (i) DDTSE \cite{zhang2024ddtse}, the DDTSE-only variant from the original paper which do not employ an additional discriminative model, and (ii) DiffSep \cite{scheibler2023diffusion}. Among discriminative approaches, we compare to (iii) DPCCN \cite{han2022dpccn}, (iv) pBSRoformer \cite{le2023personalized}, and (v) a discriminative adaptation of \cite{han2022dpccn} to NCSN++ \cite{song2020score}.
Additionally, we compare to (vi) Whisper-TSE \cite{ma2025enhancing} which combines a Whisper encoder with a flow-based tokenizer synthesizer. 
We further include two models utilizing discrete unit representations:(vii) SKIM \cite{yu2024generation}, employing the best-performing VQ-wav2vec-based discrete vocoder variant and (viii) TSELM \cite{tang2024tselm}. Results for NCSN++, DPCCN, and DiffSep are taken from \cite{zhang2024ddtse}, while pBSRoformer results are taken from \cite{ma2025enhancing}. 
We also evaluate our model on the speech enhancement setup, following \cite{zhang2024ddtse}, we compare it to the following approaches: NCSN++ \cite{song2020score}, DPCCN \cite{han2022dpccn}, SGMSE+ \cite{lemercier2023analysing}, WGSL \cite{ayilo2024diffusion}, DCCRN \cite{hu2020dccrn}, and DDTSE \cite{zhang2024ddtse}. All speech enhancement baseline results are taken from \cite{zhang2024ddtse}.

\textbf{Evaluation metrics.} We evaluate \ourmethod{} using both intrusive and non-intrusive speech quality metrics, assessing perceptual quality, intelligibility, and speaker consistency. Intrusive metrics, which require a clean reference signal, include Perceptual Evaluation of Speech Quality (PESQ)~\cite{rix2001perceptual} and Extended Short-Time Objective Intelligibility (ESTOI)~\cite{taal2010short}. 
The non-intrusive metrics, the overall quality (OVRL), and the deep noise suppression mean opinion score (DNSMOS)~\cite{reddy2022dnsmos}, provide assessments without requiring a clean reference. To measure intelligibility, we follow~\cite{ma2025enhancing} to calculate the word error rate (WER) using the Whisper-small model. Speaker similarity is evaluated using cosine similarity between speaker embeddings, which are extracted from a ResNet34 model pre-trained on VoxCeleb2 
 following \cite{zhang2024ddtse}.
Intrusive metrics such as SI-SDR 
are generally unsuitable for evaluating generative mel-based approaches~\cite{ma2025enhancing,yu2024generation,tang2024tselm}, hence are omitted from the main results. However, we provide results for these metrics using our phase-conditioned vocoder approach in Section~\ref{sec:vocoder_exp}.

\textbf{Training setup.} Following~\cite{chen2024f5}, our \ourmethod{} model consists of 22 layers, with 16 attention heads, and an embedding dimension of 1024, followed by a 2048-dimensional feed-forward network. We train our model for 100 epochs with a batch size of 11K frames using the AdamW optimizer with a peaked learning rate of 1e-4. We represent audio samples with 100-dimensional log mel-filterbank features with 24kHz sampling rate and hop length 256, for correspondence with the Vocos vocoder.

\begin{figure}[t]
\centering
\includegraphics[width=1.\linewidth]{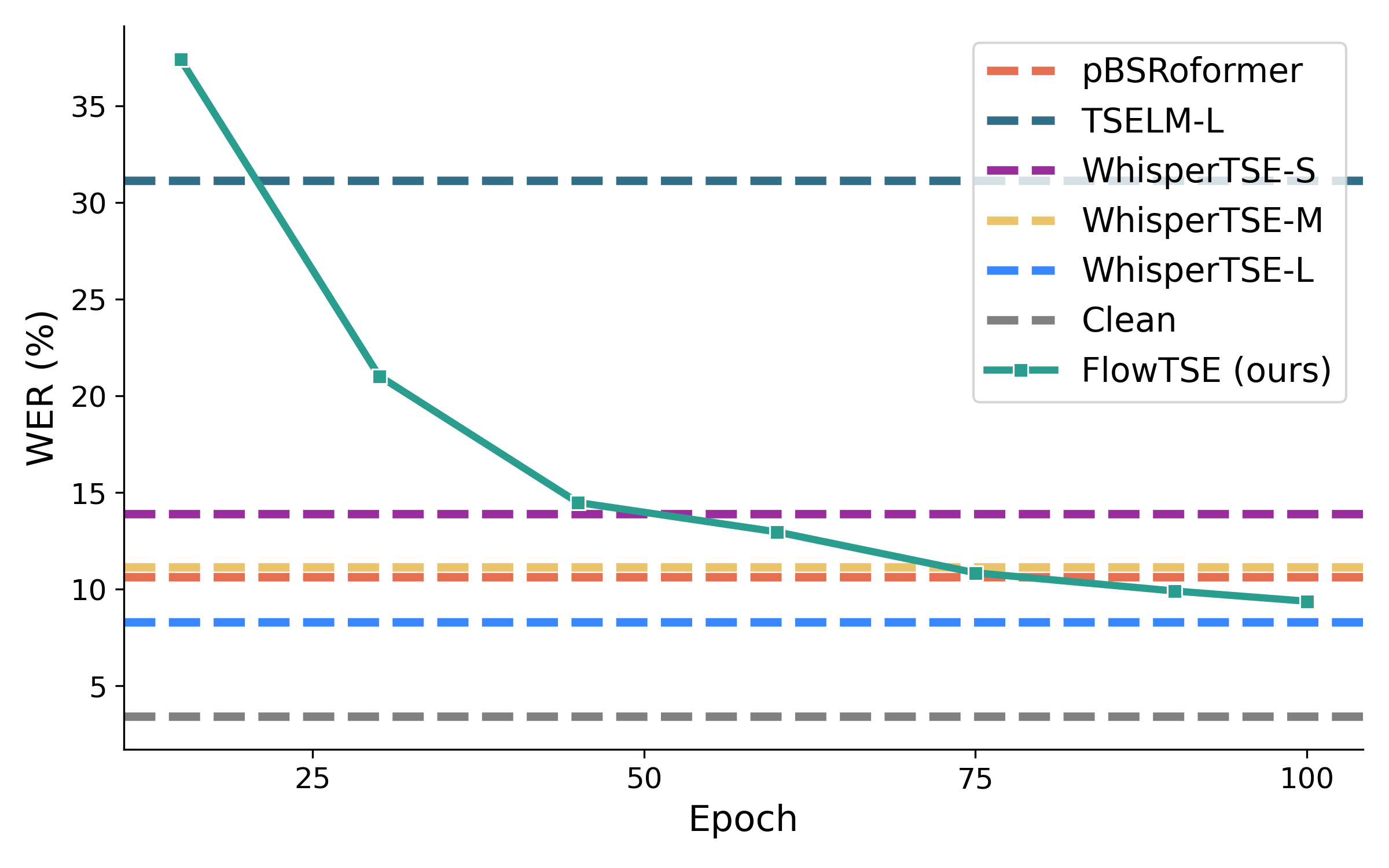}
\caption{Comparison of word error rate evaluated on the Libri2Mix clean with max split. \ourmethod{} outperforms the WhisperTSE-S and pBSRoformer baselines while achieving a WER comparable to WhisperTSE-L, despite the latter being a significantly larger model.}
\label{fig:wer}
\end{figure}

\subsection{Results}

\subsubsection{Multi-speaker TSE}\label{sec:multispeaker}
We evaluate our model under speaker mixture setups, with and without integrated noise. We use the \texttt{clean} and \texttt{both} splits of Libri2Mix under the \texttt{min} settings. The results presented in Table~\ref{tab:libri2mix} show that \ourmethod{} achieves on-par results or outperforms strong baselines, including both generative and discriminative. Importantly, \ourmethod{} achieves the highest PESQ score among all methods, demonstrating its strong performance in generating clean speech quality. Furthermore, in the \texttt{mix-clean} setting, \ourmethod{} surpasses the Clean baseline in both OVRL and DNSMOS, further showcasing its effectiveness.
Notably, it also achieves impressive SIM scores in both \texttt{mix-clean} and \texttt{mix-both} setups, highlighting its ability to maintain the speaker’s speech characteristics. 
Additionally, to better evaluate the intelligibility of our approach, we follow~\cite{ma2025enhancing} and report the word error rate evaluated using the Whisper-small model. For this experiment, we use the \texttt{max} split of the Libri2Mix mix-clean dataset, in alignment with~\cite{ma2025enhancing}. The results are presented in Figure~\ref{fig:wer}. Interestingly, our approach achieves a significantly lower WER than WhisperTSE-S, despite having a similar number of parameters as \ourmethod{}. Moreover, it attains a WER comparable with WhisperTSE-L, which is substantially larger. These baselines were directly optimized for enhanced intelligibility using target speech transcriptions as additional training supervision, by leveraging the Whisper ASR.

\subsubsection{Speech Enhancement}

Although \ourmethod{} is designed and optimized for the TSE task, it can also be applied to speech enhancement (SE). Importantly, here we use the pretrained model from Section~\ref{sec:multispeaker} with no additional training. In this scenario, the model is presented with the utterance of a single speaker in a noisy environment. To align with the standard SE setup, in this experiment we do not provide the model enrollment speech, and the model is provided with the mixed speech both as input and as the enrollment condition.
The results are presented in Table \ref{tab:single_speaker_enhancement}. As shown, \ourmethod{} achieves the best results in terms of the PESQ, OVRL and DNSMOS, with equal ESTOI to the best performing NCSN++ and DDTSE.

\begin{table}[t]
    \centering
    \caption{Performance comparison for single-speaker scenario (speech enhancement) using the \texttt{min} split of the Libri2Mix mix-single dataset.}
    \resizebox{\columnwidth}{!}{
    \begin{tabular}{l ccccc}
        \toprule
        Model & Type& PESQ & ESTOI & OVRL & DNSMOS \\
        \midrule
        Mixure & - &1.16 & 0.56 & 1.75 & 2.63 \\
        \midrule
        DCCRN & D &2.03 & 0.81 & 2.98 & 3.64 \\
        NCSN++ & D &1.85 & 0.82 & 3.11 & 3.59 \\
        \midrule
        SGMSE+ & G &1.99 & 0.82 & 3.12 & 3.60 \\
        WGSL & G &1.86 & 0.79 & 3.08 & 3.50 \\
        DDTSE & G & 2.01 & 0.82 & 3.25 & 3.75 \\
        \midrule
        \ourmethod{} 
        & G& \textbf{2.18} & \textbf{0.83} &\textbf{3.27} & \textbf{3.77} \\
        \bottomrule
    \end{tabular}
    }
    \vspace{0.1cm}
    \label{tab:single_speaker_enhancement}
\end{table}



\subsubsection{Phase-aware Vocoder}\label{sec:vocoder_exp}
We turn to evaluate the impact of phase reconstruction with our Phase-conditioned Vocoder (see \ref{sec:vocoder}). Table~\ref{tab:phase_vocoder} presents objective metrics comparing our method with and without the phase vocoder. We use the Libri2Mix (mix-clean) dataset under the \texttt{min} setup and train the vocoder for 5000 steps as outlined in Section~\ref{sec:vocoder}. Where applicable, the modules are initialized from the pretrained Vocos weights. Our phase-aware vocoder approach enhances the SI-SDR metric, as evident from the results. However, this improvement may come at the cost of lower scores in other metrics, such as PESQ, ESTOI, and OVRL. The improvement in SI-SDR is driven by accurate phase reconstruction, as this metric is highly sensitive to temporal misalignments between the generated waveform and the input speech. 


\begin{table}[h]
    \centering
    \caption{Effect of our phase-conditioned vocoder. 
    }
    \resizebox{\columnwidth}{!}{
    \begin{tabular}{lcccc}
        \toprule
        Method & PESQ & ESTOI & SI-SDR & OVRL \\
        \midrule
        Without Phase Vocoder &  \textbf{2.58} & \textbf{0.84} & -25.64 & \textbf{3.27} \\
        With Phase Vocoder & 2.48 & \textbf{0.84} & \textbf{10.81} & 3.14 \\
        \bottomrule
    \end{tabular}
    }
    \label{tab:phase_vocoder}
\end{table}

\section{Discussion}
In this work, we introduce \ourmethod{}, a novel generative approach for high-quality TSE. We leverage recent advancements in flow matching to condition our model on the target speaker’s audio. This simple yet effective approach reduces computational overhead and eliminates the need for pretrained models in complex pipelines.
Through extensive experiments, we demonstrate that despite its simplicity, \ourmethod{} produces audio with high intelligibility and naturalness, matching or surpassing recent TSE baselines.
We believe that \ourmethod{} drives TSE toward more simplified and effective generative approaches.
\bibliographystyle{IEEEtran}
\bibliography{main}

\end{document}